\documentclass[prl,nofootinbib,superscriptaddress,twocolumn]{revtex4-1}
\usepackage[a4paper,left=1.5cm,right=1.5cm,top=3cm,bottom=3cm]{geometry}

\usepackage{amssymb,amsmath,amsfonts}
\usepackage[dvipsnames]{xcolor}
\usepackage{graphicx}
\usepackage{longtable}
\usepackage{verbatim}
\usepackage{color}
\usepackage{mdframed}
\usepackage{soul}
\usepackage{amsfonts,amssymb,mathrsfs,amsmath}
\usepackage{slashed, cancel}
\usepackage{framed}
\usepackage{mdframed}
\usepackage{simplewick} 
\allowdisplaybreaks

\usepackage{latexsym}
\usepackage{graphicx}
\usepackage[dvipsnames]{xcolor}
\usepackage{booktabs}
\usepackage{datetime}
\newdateformat{mydate}{\THEDAY{ }\monthname[\THEMONTH]{ }\THEYEAR}

\usepackage{tikz}
\usepackage{color}
\usepackage{framed}
\usepackage{hyperref}
\hypersetup{colorlinks, citecolor=bluscuro, linkcolor=black, urlcolor=bluscuro}
\definecolor{rossos}{cmyk}{0,1,1,0.55}
\definecolor{bluscuro}{rgb}{0.15, 0.2, .85}
\definecolor{bluchiaro}{cmyk}{1,.3,0.,0.1}

\newcommand{\be}{\begin{equation}}
\newcommand{\ee}{\end{equation}}
\newcommand{\bea}{\begin{eqnarray}}
\newcommand{\eea}{\end{eqnarray}}
\newcommand{\beq}{\begin{equation}}
\newcommand{\eeq}{\end{equation}}
\newcommand{\lp}{\left(}
\newcommand{\rp}{\right)}
\newcommand{\llp}{\left[}
\newcommand{\rrp}{\right]}

\def\beqa{\begin{eqnarray}}

\def\f{{\text{\tiny f}}}
\def\ii{{\text{\tiny i}}}

\def\e{\text{\tiny e}}
\def\d{{\rm d}}

\def\cutoff{\text{\tiny cut-off}}

\def\PBH{\text{\tiny PBH}}

\def\eeqa{\end{eqnarray}}

\def\lsim{\mathrel{\rlap{\lower4pt\hbox{\hskip0.5pt$\sim$}}
    \raise1pt\hbox{$<$}}}         
\def\gsim{\mathrel{\rlap{\lower4pt\hbox{\hskip0.5pt$\sim$}}
    \raise1pt\hbox{$>$}}}         

\usepackage[normalem]{ulem}

\newcommand{\arXiv}[2]{\href{http://arxiv.org/pdf/#1}{{\tt [#2/#1]}}}
\newcommand{\arXivold}[1]{\href{http://arxiv.org/pdf/#1}{{\tt [#1]}}}

\begin{document}

\title{Constraints on Primordial Black Holes:
the Importance of Accretion}

\author{V. De Luca}
\address{D\'epartement de Physique Th\'eorique and Centre for Astroparticle Physics (CAP), Universit\'e de Gen\`eve, 24 quai E. Ansermet, CH-1211 Geneva, Switzerland}

\author{G. Franciolini}
\address{D\'epartement de Physique Th\'eorique and Centre for Astroparticle Physics (CAP), Universit\'e de Gen\`eve, 24 quai E. Ansermet, CH-1211 Geneva, Switzerland}
\address{Instituto de F\'isica Te\'orica UAM-CSIC, Universidad Aut\'onoma de Madrid, Cantoblanco, Madrid, 28049 Spain}

\author{P. Pani}
\address{Dipartimento di Fisica, “Sapienza” Università di Roma, Piazzale Aldo Moro 5, 00185, Roma, Italy}
\address{INFN, Sezione di Roma, Piazzale Aldo Moro 2, 00185, Roma, Italy}

\author{A.~Riotto}
\address{D\'epartement de Physique Th\'eorique and Centre for Astroparticle Physics (CAP), Universit\'e de Gen\`eve, 24 quai E. Ansermet, CH-1211 Geneva, Switzerland}

\address{INFN, Sezione di Roma, Piazzale Aldo Moro 2, 00185, Roma, Italy}

\date{\today}

\begin{abstract}
\noindent
We consider the constraints on the fraction of  dark matter in the universe in the form of primordial black holes  taking into 
account the crucial role of accretion which may change  both their
mass and mass function. We show that accretion may drastically weaken the  constraints at the present epoch 
for primordial black holes with masses larger than a few solar masses.
\end{abstract}

\maketitle

\paragraph{Introduction.}
\noindent
The detection of gravitational waves generated by the merger of black holes~\cite{Abbott:2016blz, 
TheLIGOScientific:2016pea, LIGOScientific:2018jsj, LIGOScientific:2018mvr}  has given rise to a 
renewed interest in the idea that primordial black holes~(PBHs) could compose a fraction (or all) of the dark 
matter~(DM)~\cite{Bird:2016dcv} (see Refs.~\cite{sasaki, revPBH1,Carr:2020gox} for reviews).  
 
In order to firmly  assess if  PBHs with a given mass might contribute to a significant fraction of the DM,  a 
careful investigation should be performed to understand whether the current observational constraints on the PBH 
abundance --~usually parametrised by the fraction $f_\PBH\equiv \Omega_\PBH/\Omega_{\text{\tiny DM}}$  at the present 
epoch~\cite{Carr:2020gox}~-- apply around that mass.
 
It goes without saying that understanding what physical phenomena   can alter such constraints, either  strengthening 
or weakening them, is of utmost importance.   For instance, a large spatial clustering of PBHs might  in principle help 
in avoiding microlensing constraints~\cite{gb}, even though one should take into account that PBHs are initially not
clustered (in the absence of primordial  non-Gaussianity)~\cite{cl1,cl2,cl3,dizgah}. 
 
PBH mergers and accretion can also have an impact on the PBH bounds. 
While  only a tiny fraction of PBHs detectable through their coalescence have experienced  a previous 
merger event~\cite{us}, PBHs 
may efficiently accrete during the cosmic history~\cite{Ricotti:2007jk}. In particular, if they do not represent the 
only  DM component, PBHs may accrete a DM halo, thus increasing  their gravitational potential and  the ordinary gas 
accretion~\cite{Ricotti:2007au}.

For sufficiently massive PBHs, accretion may occur at super-Eddington rates up to the reionization 
epoch~\cite{Ricotti:2007jk,Ricotti:2007au,us}. 
The mass distribution of PBHs at low redshift  is therefore  different from the one  at high redshift. This implies that 
limits on the \emph{current} PBH abundance, which are expressed in terms of the \emph{present} PBH mass values,
must be properly revisited. For example, the CMB 
temperature and polarization fluctuations are sensitive to the energy injection up to redshift $z\sim 300$~\cite{Ricotti:2007au,serpico}. 
In the presence of significant accretion up to much lower redshifts, the 
PBHs masses and their distribution measured today (albeit indirectly) do not coincide with those at redshifts 
$z\sim (300\div 600)$ when the energy deposition has the largest effect on the CMB. 

The goal of this paper is to highlight the significant role played by the phenomenon of accretion in setting the 
observational limits on the current PBHs abundance. We therefore proceed to briefly describe the main features of 
accretion onto PBHs and then investigate its impact on the current constraints on $f_\PBH$. 

\vskip 0.3cm
\noindent
\paragraph{Accretion onto PBHs.}
\noindent
Once PBHs are produced, for instance due to the collapse of sizable overdensities in the radiation-dominated epoch, 
their evolution may be significantly affected by accretion during the cosmic history~\cite{Ricotti:2007au,zhang}. 
The physics of accretion is very complex, since the accretion rate and the geometry of the accretion flow are 
intertwined, and they are both crucial in determining the evolution of the PBH mass. 

A PBH  of a given mass $M$ immersed in the intergalactic medium can accrete baryonic matter at the Bondi-Hoyle rate
given by~\cite{ShapiroTeukolsky}
\be
\label{R1}
\dot{M}_\text{\tiny B} = 4 \pi \lambda m_H n_{\rm gas} v_\text{\tiny eff} r_\text{\tiny B}^2\,,
\ee
where $r_\text{\tiny B} = GM/v^2_\text{\tiny eff}$ is the Bondi-Hoyle radius, $n_{\rm gas}$ is the hydrogen gas number 
density, $v_\text{\tiny eff} = \sqrt{v^2_\text{\tiny rel} + c_s^2}$ is the PBH effective velocity, defined in terms of 
the PBH relative velocity $v_\text{\tiny rel}$ with respect to the gas with sound speed $c_s$. The accretion parameter 
$\lambda$ takes into account the gas viscosity, the Hubble expansion, and the coupling of the gas to the CMB radiation 
through Compton scattering.  Its explicit expression depends on redshift, ionisation fraction $x_e$, PBH mass and 
effective velocity, and is given by \cite{Ricotti:2007jk, Ricotti:2007au,us}
\be
\lambda = {\rm exp} \lp \frac{9/2}{3 + \hat{\beta}^{0.75}} \rp x_{\rm cr}^2,
\ee
in terms of the gas viscosity parameter 
\begin{align}
\hat{\beta} &= \lp \frac{M}{10^4 M_\odot} \rp \lp \frac{1+z}{1000}\rp^{3/2} \lp \frac{v_\text{\tiny eff}}{5.74 \, {\rm 
		km \, s^{-1}}} \rp^{-3} \nonumber \\
& \times \llp 0.257 + 1.45 \lp \frac{x_e}{0.01}\rp \lp \frac{1+z}{1000}\rp^{5/2} \rrp,
\end{align}
and the sonic radius 
\be
x_{\rm cr} \equiv \frac{r_{\rm cr}}{r_\text{\tiny B}}= \frac{-1 + (1+ \hat{\beta})^{1/2}}{ \hat{\beta}}.
\ee
If PBHs do not comprise the totality of DM in the universe, one has to consider the additional presence of a dominant DM 
component which forms, around each PBH, a dark halo of mass~\cite{Mack:2006gz, Adamek:2019gns}
\be
M_h(z) = 3M \left(\frac{1+z}{1000} \right)^{-1}\,,
\ee
which grows with time as long as the PBH does not interact with others.
While direct accretion of DM is negligible for the PBH evolution~\cite{Ricotti:2007au}, the effect of this DM 
clothing is to enhance the gas accretion rate, acting in this way as a catalyst. 
The characteristic halo radius, assuming a power law density profile $\rho \sim r^{-\alpha}$ \cite{Ricotti:2007au,Adamek:2019gns,Bertschinger:1985pd}, is given by 
\be
r_h = 0.019 \, {\rm pc} \left( \frac{M}{M_\odot} \right)^{1/3}\left(\frac{1+z}{1000} \right)^{-1},
\ee
and has to be confronted with the Bondi radius by introducing the parameter
\be
\kappa \equiv \frac{r_\text{\tiny B}}{r_h} = 0.22\lp \frac{1+z}{1000}\rp \lp \frac{M_h}{M_\odot}\rp^{2/3} \lp 
\frac{v_\text{\tiny eff}}{{\rm km \, s^{-1}}} \rp^{-2}\,.
\ee
In the regime $\kappa\geq2$ the dark halo behaves as a point mass and the accretion rate will be the same as one for a 
PBH of mass $M_h$. In the opposite regime, $\kappa <2$, one can introduce corrections to the naked case by modifying 
the accretion parameter as~\cite{Mack:2006gz}
\be
\hat{\beta}^{h} \equiv \kappa^{\frac{p}{1-p}} \hat{\beta}, \quad \lambda^{h} \equiv \bar\Upsilon^{\frac{p}{1-p}} 
\lambda 
(\hat{\beta}^{h}), \quad r_{\rm cr}^h \equiv \lp \frac{\kappa}{2} \rp^{\frac{p}{1-p}} r_{\rm cr},
\ee
where $p = 2- \alpha$ and
\be
\bar\Upsilon = \lp 1 + 10 \hat{\beta}^h \rp^{\frac{1}{10}} {\rm exp} (2 - \kappa) \lp \frac{\kappa}{2} \rp^2.
\ee
One can normalise the Bondi-Hoyle accretion rate in terms of the Eddington one
\be
\dot M_\text{\tiny Edd} = 1.44 \cdot 10^{17} \, \lp \frac{M}{M_\odot} \rp {\rm g}  \, {\rm s}^{-1}
\ee
by introducing the dimensionless rate
\be
\dot{m} \equiv \frac{\dot{M}_\text{\tiny B}}{\dot{M}_\text{\tiny Edd}} = 0.023 \lambda \lp \frac{1+z}{1000}\rp \lp 
\frac{M}{M_\odot}\rp \lp \frac{v_\text{\tiny eff}}{5.74 \, {\rm km \, s^{-1}}} \rp^{-3},
\ee
such that the mass evolution equation takes the more compact form (see, e.g., Refs.~\cite{Barausse:2014tra,us})
\be
\dot M \sim 0.002\, \dot m(M) \left(\frac{M}{10^6 M_\odot} \right) M_\odot \,{\rm yr}^{-1}.
\label{Mdot}
\ee
The dimensionless baryonic accretion rate $\dot m$ carries all the 
information on the geometry and efficiency of the accretion process. For example, if the angular momentum carried by the baryonic infalling material is 
sufficiently high, a thin accretion disk forms around the PBH.
Details on that can be found in Ref.~\cite{us} and references therein.

\begin{figure}[t!]
	\includegraphics[width=\columnwidth]{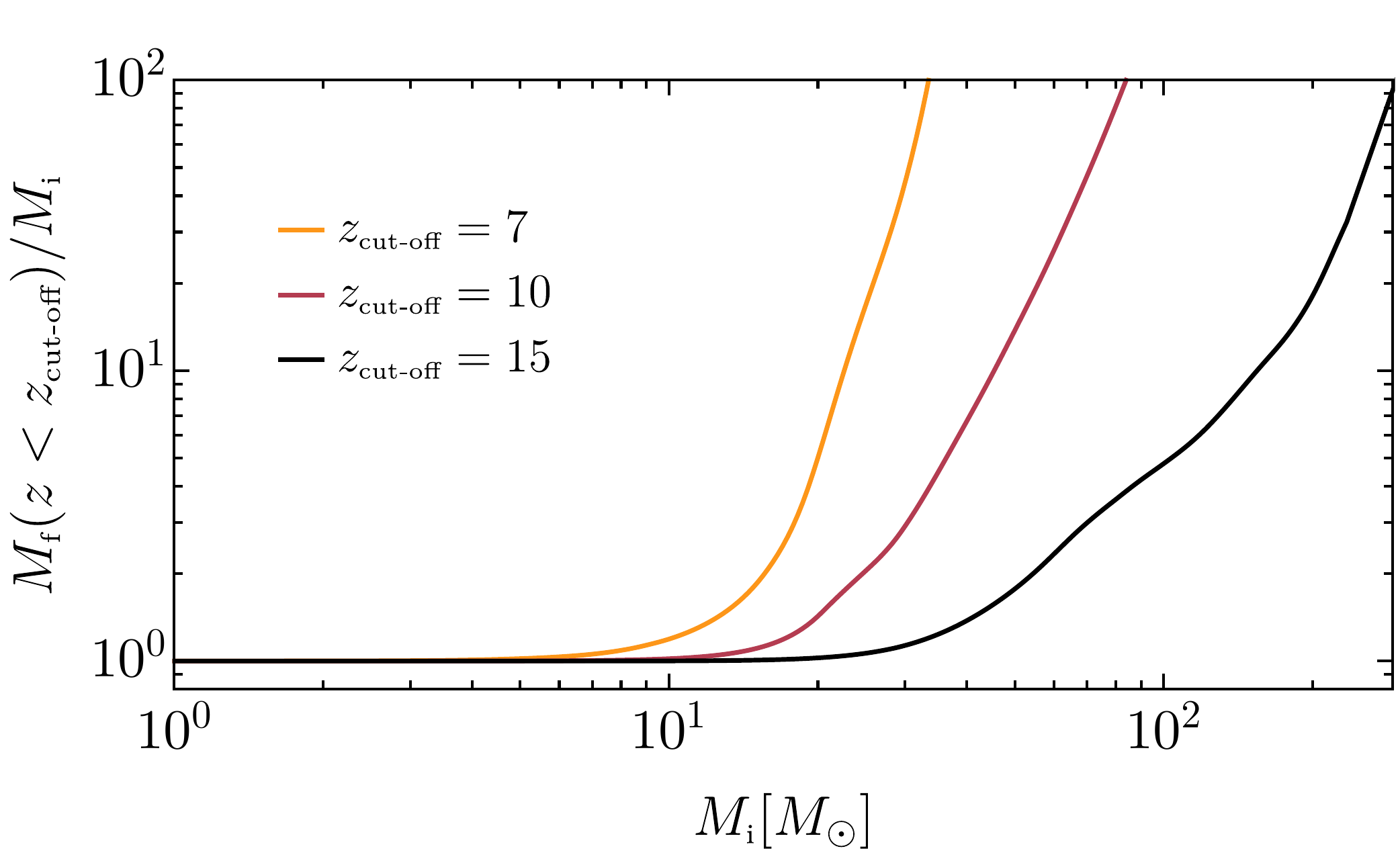}
	\caption{Final mass $M_\text{\tiny f}$ at a given redshift $z$ obtained from the evolution of an initial mass 
$M_\text{\tiny i}$ in the various scenarios with cut-off redshift $z_\text{\tiny cut-off}= 15$, $10$ and $7$.
}
	\label{fig1M}
\end{figure}

From Eq.~\eqref{Mdot}, the typical accretion timescale is $\tau_\text{\tiny ACC} = \tau_\text{\tiny Salp}/ \dot 
m$, where $\tau_\text{\tiny Salp} = 4.5 \times 10^8 \,{\rm yr}$ is the  Salpeter time. 
For redshifts $z \gsim 30$ the accretion timescale is much longer than the age of the universe at that epoch, and 
therefore there is not enough time for the accretion process to play a significant role in the PBH mass evolution. 
Depending on the PBH masses, after $z\sim 30$ accretion may have an important impact, reaching also super-Eddington 
values (see Fig.~4 in Ref.~\cite{us}).

At smaller redshifts, uncertainties in the accretion model come into play. While local feedbacks are not relevant for the  range of PBH masses of our interest~\cite{Ricotti:2007au}, global feedbacks from the radiation emitted by the gas accreting onto the PBHs, along with pre-heating effects from X-ray backgrounds~\cite{Oh:2003pm}, may heat and ionize the intergalactic medium, leading to sensible changes in its temperature.
Moreover, with the beginning of structure formation, one expects the fall of a large part of the PBH population into the 
gravitational potential wells of the large-scale structures, leading to an increase of the PBH peculiar velocities, see in particular  Ref.~\cite{Hasinger:2020ptw} for a recent analysis. 
This effect, together with reionization and global feedbacks, may lead to a decrease of the accretion 
rate~\cite{Ricotti:2007au,Ali-Haimoud:2016mbv, Ali-Haimoud:2017rtz, raidalsm,Inman:2019wvr}. 
Owing to the uncertainties in the modeling of accretion onto PBHs at redshift $z\sim 10$, we have decided to consider three different cut-off points and assume that mass accretion becomes 
negligible after at $z_\cutoff\simeq 15$, $10$ and $7$.  The first value has been chosen to show a moderate effect of 
accretion, the second value corresponds to Model~I of Ref.~\cite{us} and we consider it the most realistic choice, 
whereas the third value considers the case in which accretion onto PBHs is significant at a relatively smaller redshift, 
as discussed in Ref.~\cite{Ricotti:2007au}.

The accretion-driven evolution of the mass is shown in Fig.~\ref{fig1M}. For initial masses $M_\ii$ smaller than 
a few solar masses, depending on $z_\cutoff$, accretion does not have a noticeable 
impact on the PBHs mass, which remains the same up to present time. For higher masses, instead, the accretion plays an 
important role, leading to a final mass $M_\f$ at the present epoch which can be various orders of magnitude larger than 
the initial one.

\begin{figure}[t!]
	\includegraphics[width=\columnwidth]{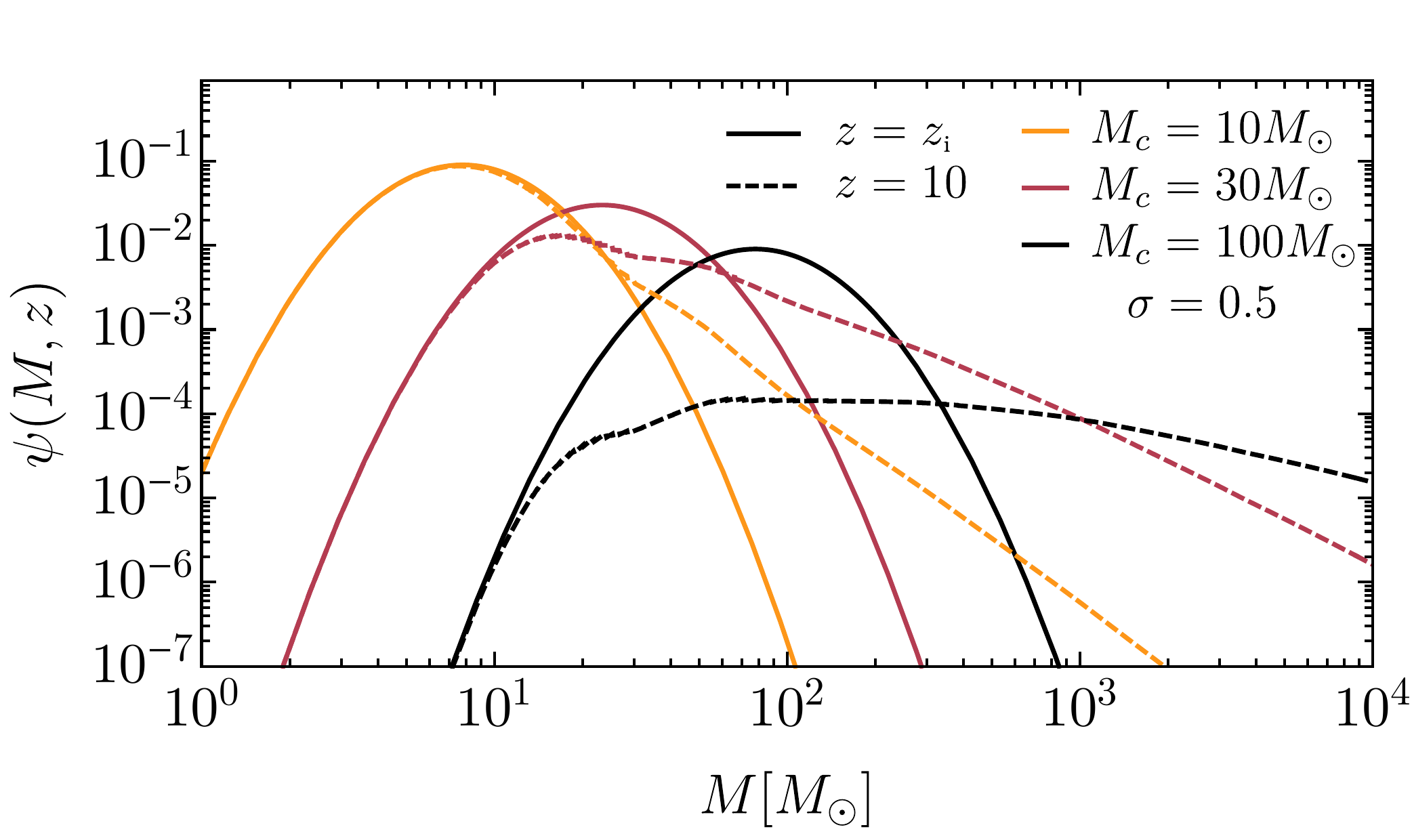}
	\caption{Evolution of the mass function from formation (solid lines) to redshift $z = 10$ (dashed lines), for an 
initial lognormal shape with width $\sigma = 0.5$ at different values of $M_c$.}
	\label{figmassfunct}
\end{figure}

Accretion has also the additional effect of changing the PBH mass distribution with redshift. We define the mass function $\psi(M,z)$ as the fraction of PBHs with mass in the interval $(M, M + \d M)$ at redshift $z$. For an 
initial $\psi (M,z_\ii)$ at formation redshift $z_\ii$, its evolution is governed by~\cite{us}
\begin{equation}
	\psi(M_\text{\tiny f}(M,z),z) \d M_\text{\tiny f} = \psi(M,z_\ii) \d M \label{psiev}
\end{equation}
where $M_\text{\tiny f}(M,z)$ is given in Fig.~\ref{fig1M}.
When accretion is present, also the value of $f_\PBH$ depends on the redshift as\footnote{
We assume here for simplicity a non-relativistic dominant DM component, whose energy density scales as the inverse of the volume.}
\begin{align}
\label{fev}
f_\PBH(z)  &= \frac{\rho_\PBH}{(\rho_\text{\tiny DM} - \rho_\PBH) + \rho_\PBH} \nonumber \\
& = \frac{\langle M(z)\rangle}{\langle M(z_\ii)\rangle(f^{-1}_\PBH(z_\ii)-1)+\langle M(z)\rangle},
\end{align}
 defined in terms of the average mass 
\begin{equation}
\langle M(z)\rangle =\int \d M M \psi (M, z).
\end{equation}
One can consider several initial shapes for the mass 
function at high redshift, depending on the details of the formation mechanism. 
Often in the literature an initial sharp 
monochromatic population is considered, with a constant reference peak mass $M_c$. Accretion has the effect of shifting such a peak. 
For more physically motivated scenarios, one can consider an initial extended mass function given by a lognormal shape with 
width $\sigma$,
\be
\label{psi}
\psi (M,z_\ii) = \frac{1}{\sqrt{2 \pi} \sigma M} {\rm exp} \left(-\frac{{\rm log}^2(M/M_c)}{2 \sigma^2} \right)\,,
\ee
 properly normalised to unity and whose evolution with time is shown in Fig.~\ref{figmassfunct}. 
As one can see, as $M_c$ increases, the evolved mass function becomes increasingly flatter with a high-mass tail 
orders of magnitude above its corresponding value at formation. 

\begin{figure}[t!]
	\includegraphics[width=\columnwidth]{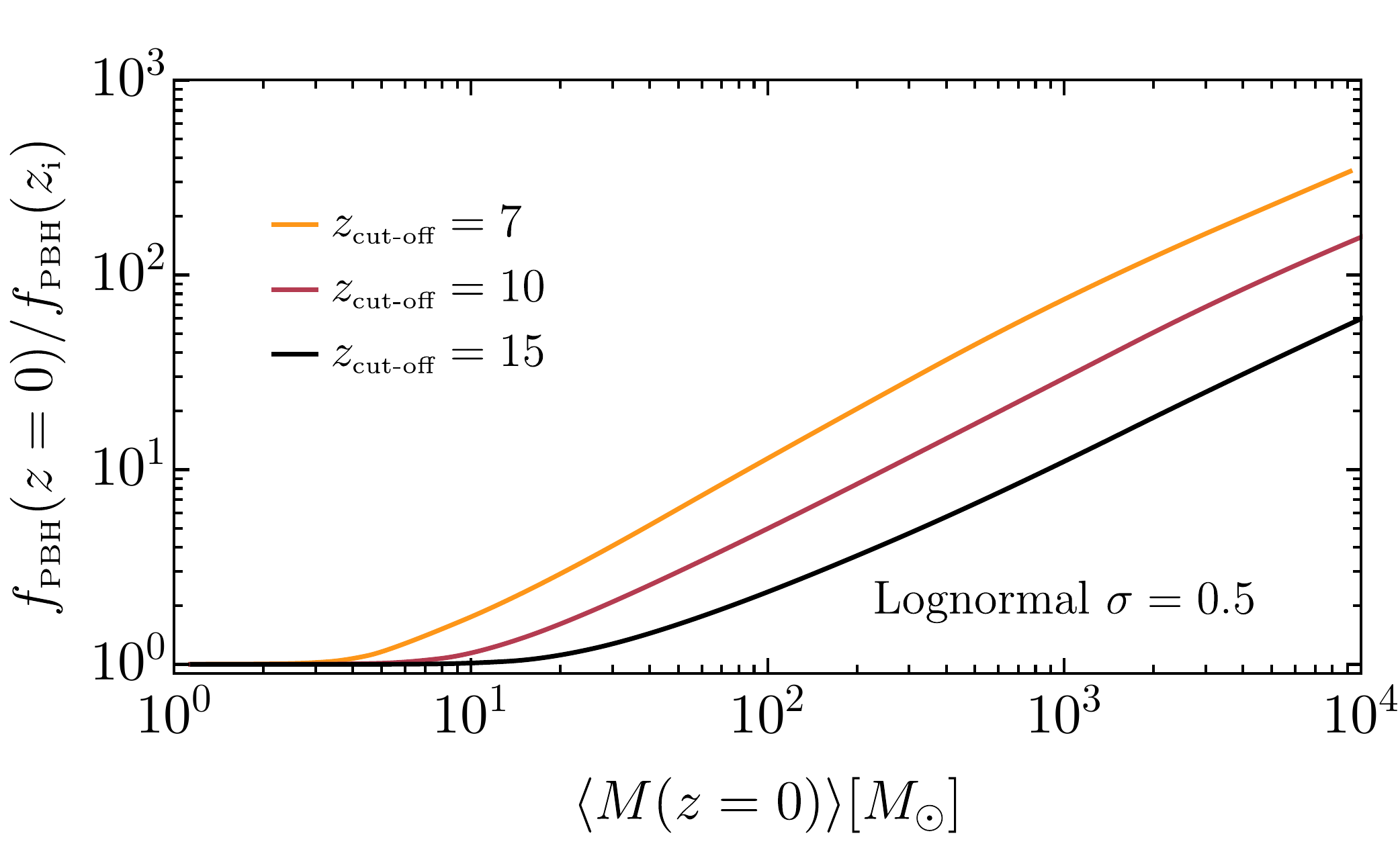}
	\caption{Ratio between the initial and late time PBH abundance depending on $\langle M (z=0)\rangle$ for an initial 
lognormal mass function with $\sigma = 0.5$, away from the saturation happening when $f_\PBH$ approaches unity.}
	\label{figmassfunct-2}
\end{figure}

The evolution of the abundance $f_\PBH$ in the case of a monochromatic mass function can be read off Fig.~\ref{fig1M}, 
as $f_\PBH(z)/f_\PBH(z_\ii)\sim M_\f(z)/M_\ii$ as long as the abundance is not close to unity. In a similar manner, the 
corresponding  evolution in the case of a lognormal mass function is plotted in Fig.~\ref{figmassfunct-2} as a function 
of $\langle M(z=0)\rangle$.

\vskip 0.3cm
\noindent
\paragraph{Results.}
\noindent
Detailed investigations on the current observational constraints on the PBH abundance have been performed in the 
literature (see, e.g., Ref.~\cite{Carr:2020gox} for a recent review on the topic). In the range of masses 
affected 
by accretion, the most important constraints come from lensing, dynamical processes, formation of structures, and 
accretion related phenomena. The lensing bounds include those from Supernovae ~\cite{Zumalacarregui:2017qqd}, the 
MACHO and EROS experiments~\cite{Alcock:2000kd, Allsman:2000kg}, ICARUS (I)~\cite{Oguri:2017ock} and radio 
~\cite{Wilkinson:2001vv} observations. They all consider lensing sources at low redshift $z \ll z_\cutoff$. Dynamical 
constraints involve disruption of wide binaries~\cite{Quinn:2009zg}, and survival of star clusters in Eridanus II~\cite{Brandt:2016aco} and Segue I ~\cite{Koushiappas:2017chw} at small redshifts. Bounds also arise by 
observations of the Lyman-$\alpha$ forest at redshift before $z \approx 4$~\cite{Murgia:2019duy}. Other 
constraints involve bounds from Planck data on the CMB anisotropies induced by X-rays emitted by spherical or 
disk (Planck~D)~\cite{Ali-Haimoud:2016mbv, serpico} accretion at high redshifts or bounds on the observed 
number of X-ray (Xr) \cite{Gaggero:2016dpq, Manshanden:2018tze} and X-ray binaries (XrB) at low redshifts~\cite{Inoue:2017csr}. Additional constraints on the primordial abundance can also be 
set by the LIGO-Virgo observations~\cite{Authors:2019qbw}. 
A comprehensive plot of all the constraints is shown in Fig.~10 of Ref.~\cite{Carr:2020gox} for the choice of a 
monochromatic mass function, while in the same reference the extension to broader mass functions has been obtained 
following the procedures outlined in Refs.~\cite{Carr:2017jsz, bellomo}.

However, these constraints are standardly expressed in terms of the maximum fraction\footnote{ The constraints are in fact sensitive to 
the PBH number density $n_\PBH$ and, when translating the bounds in terms of $f_\PBH$, one has to be careful in 
properly accounting for the accretion effects on an experiment-by-experiment basis. } $f_\PBH$ of PBHs allowed in a given 
mass range; these quantities refer to the \emph{present epoch}, without taking into account the evolution of the 
PBH mass due to accretion, whose effects on the constraints are presented in the following.
Motivated by the fact that the current LIGO-Virgo observations are at low redshifts, we have decided to present our 
results in terms of the values of $f_\PBH(z=0)$ and $ \langle M(z=0)\rangle$ today. Of course, for future data, for 
instance from the Einstein Telescope~\cite{Hild:2010id}, the corresponding values can be easily evaluated for a given 
initial mass function at the redshift of interest.

For simplicity, for a given PBH mass range, we consider only the most stringent constraint.
Following the prescription described in Ref.~\cite{Carr:2017jsz}, we estimate the bound on the fraction $f_\PBH 
(z_\e)$ of PBHs as DM at the redshift $z_\e$ of a given experiment from 
\begin{equation}
f_\PBH (z_\e) \lesssim  \left( \int_{M_{\rm min}(z_\e)}^{M_{\rm max}(z_\e)} \d M 
\frac{\psi (M,z_\e)}{f_{\rm max} (M,z_\e)}\right) ^{-1},
\end{equation}
where $M_\text{\tiny min}(z_\e)$ and $M_\text{\tiny max}(z_\e)$ identify the range of masses affected by the given 
constraint, and $f_\text{\tiny max} (M, z_\e)$ represents the maximum allowed fraction for a monochromatic mass 
function at the redshift of the experiment~\cite{Carr:2017jsz,Carr:2020gox}.

\begin{figure*}[t!]
	\includegraphics[width=0.48\linewidth]{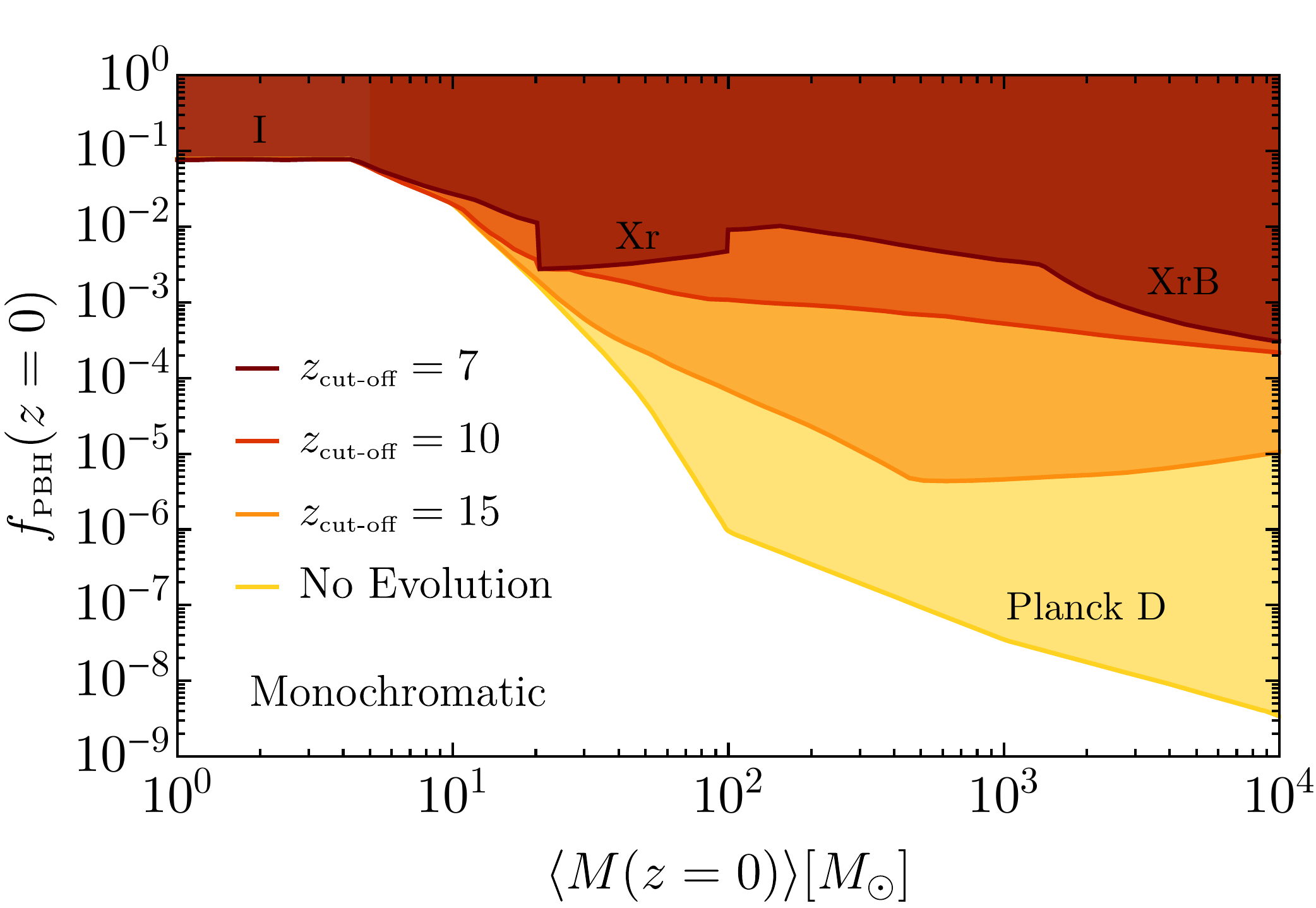}
		\includegraphics[width=0.48\linewidth]{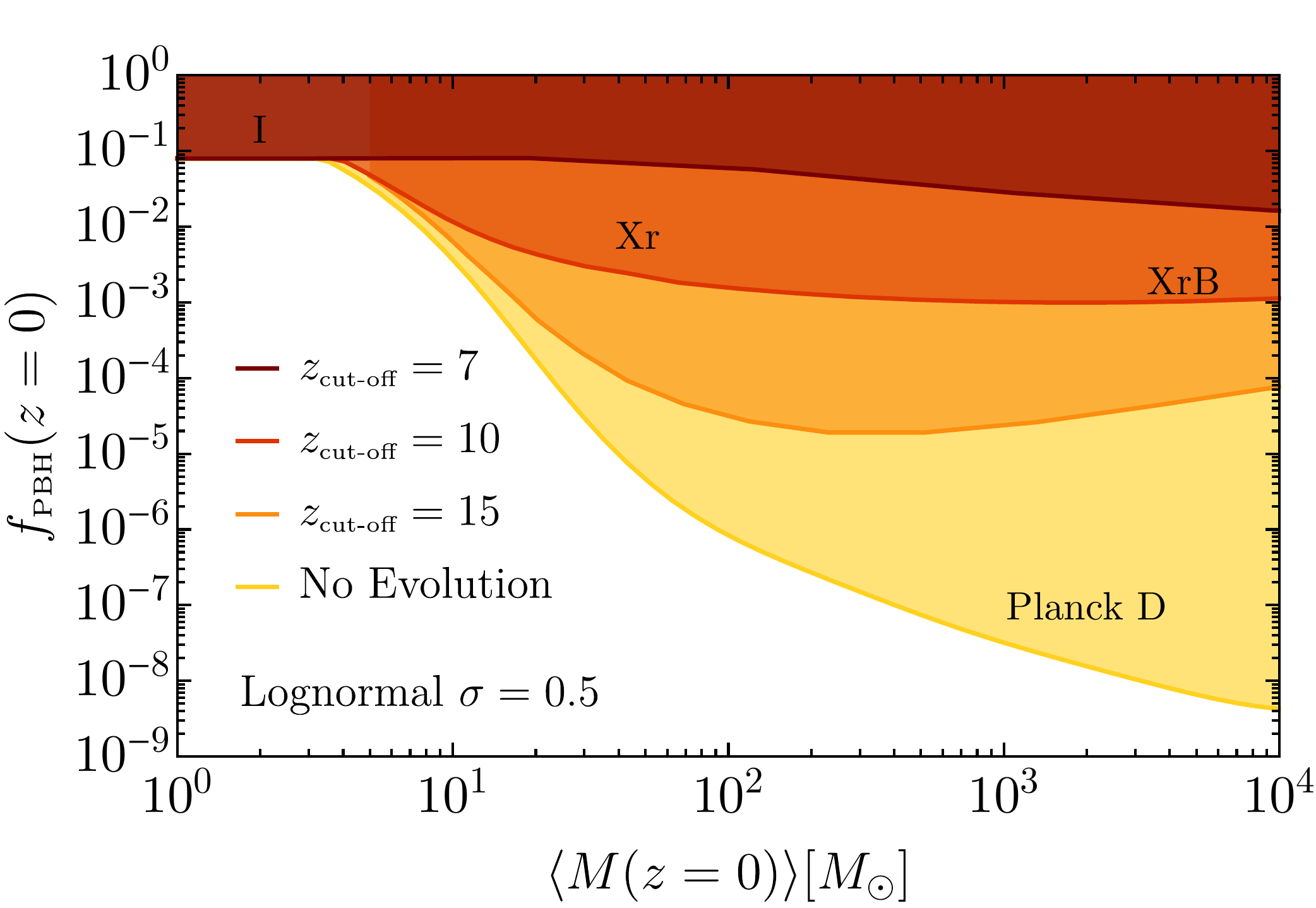}
	\caption{Combined constraints on the fraction of DM composed by PBHs today in terms of $\langle M(z = 0) 
\rangle$ for the different accretion models 
corresponding to $z_\cutoff = 15$, $10$ and $7$ compared to the case in which accretion is neglected (``No 
Evolution''). Left: monochromatic case. Right: lognormal mass function with width $\sigma =0.5$ at formation.}
	\label{figs}
\end{figure*}

Thus, for a given constraint $f_{\rm max} (M,z_\e)$ obtained neglecting accretion, we can compute the above integral by 
taking into account the evolution of the mass function from formation to $z_\e$ through Eq.~\eqref{psiev}. Finally, the 
bound $f_\PBH (z_\e)$ is mapped to $f_\PBH (z=0)$ using Eq.~\eqref{fev}. With this procedure
one gets the bound on the present fraction of PBHs as DM $f_\PBH(z=0)$ for a given $\langle M(z=0)\rangle$ as shown in 
Fig.~\ref{figs} for a monochromatic (left panel) and a lognormal (right panel) mass function. The various values of 
$\langle M(z=0)\rangle$ have been calculated by varying $M_c$.

We choose to plot a single envelope resulting from the most stringent bound for any value of $\langle M(z=0)\rangle$, 
while the labels identify the corresponding experiment dominating each portion of the graphs. Also, we compare the 
results in the scenarios with $z_\text{\tiny cut-off}=15$, $10$ and $7$ with the original constraints which neglect the 
effect of accretion.
It is worthwhile to stress that the bound Planck~D from 
the CMB is characterized by $z_\e\gsim 300$ and therefore has to be computed at that redshift using the initial mass function, as 
standardly done in the literature. 
As for the other experiments, for which lower values of $z_\e$ are involved, the bounds 
have to be estimated with the evolved mass function. 

As one can appreciate from Fig.~\ref{figs}, the observational bounds are drastically weakened  at the 
present epoch, which is of particular importance when asking the question whether a given merger event is consistent 
with the hypothesis that the black holes are of primordial origin. 
In particular, the relaxation of the bounds for a monochromatic mass function depends solely on the shift of the peak 
of the distribution and the corresponding evolution of $f_\PBH(z)$. 
 In this case, since the bounds in Fig.~\ref{figs} refer to the average PBH mass at the present epoch, only 
the constraints obtained with observations at high redshifts are affected (i.e. Planck~D in the left panel of 
Fig.~\ref{figs}).
Note that, even if the constraints obtained with observations at small 
redshifts are unaffected in the case of a monochromatic distribution, they still refer to a different mass $\langle M(z=0)\rangle$ 
relative to the case in which accretion is absent, for which $\langle M(z=0)\rangle=M_c$.  Therefore, 
if one wants to constrain the PBH formation scenario when accretion is present, and in particular the parameter $M_c$, one needs to use the mapping between $\langle M(z=0)\rangle$ and $M_c$.
On the other hand, for extended distributions (right panel of Fig.~\ref{figs}), the relaxation of the constraints 
depends also on the more complex evolution of $\psi(M,z)$, which affects not only the bounds at high redshifts, but 
also those at low redshifts, since the latter are inferred assuming a broader distribution.

\noindent
\vskip 0.3cm
\paragraph{Conclusions.}
\noindent
We have described how accretion onto PBHs may change the interpretation of the observational bounds on the current 
fraction of PBHs in DM for a given mass range. Our goal was not to
perform a comprehensive study, but just to show the crucial impact of PBH accretion on current bounds. 
We have assumed an accretion model valid for isolated PBHs. If the latter
form a binary at $z\gsim 30$~\cite{sasaki}, the effects of the binary for
the mass accretion rate should be taken into account~\cite{us}. Since
the overall fraction of PBH in binaries is  $\sim 10^{-2} f_\PBH^{16/37}$~\cite{aa}, this
effect is negligible for our bounds, except possibly for those coming
from mergers~\cite{DeLuca:2020qqa}.

Our findings are relevant in the context of the origin of the black hole mergers observed by current gravitational-wave 
interferometers. Indeed, in the mass range $(10 \div 100)M_\odot$ accretion can uplift the existing upper limits on 
$f_\PBH(z=0)$ by several orders of magnitude. The effect of accretion on PBHs is intrinsically redshift dependent, so 
it would be interesting to investigate the consequences of our results for the forecasts of future experiments, 
like the Einstein Telescope, which will probe higher redshifts and higher PBH masses.
It would be also interesting to fully assess the impact of accretion  onto the PBH merger rate and on the corresponding bounds from LIGO-Virgo observations \cite{Authors:2019qbw} when accounting for DM clothing. 
We will investigate this issue in a separate publication 
\cite{DeLuca:2020qqa}.

Our analysis can be improved along several ways. One is certainly having a better
knowledge of the impact of large-scale structures, reionization and global thermal feedbacks onto the PBH accretion; 
another one is a better characterization of accretion for values of the PBH fraction close to unity, where our 
assumption of DM clothing ceases to be correct. This would be particularly relevant in order to understand the fate of 
$f_\PBH$ in the case of strong accretion, since as one can see from Eq.~\eqref{fev}, when $\langle M(z)\rangle\gg \langle 
M(z_\ii)\rangle/f_\PBH(z_\ii)$, $f_\PBH(z)$ can dynamically approach unity, even if the initial fraction of PBHs in DM 
is 
negligible. 
Finally, a more detailed characterization of accretion would be instrumental to assess if relatively light PBHs may 
play a role in explaining the supermassive black holes observed at $z\gsim 6$~\cite{Inayoshi:2019fun}.

\vskip 0.3cm
\noindent
\paragraph{Acknowledgments.}
\noindent
We thank B.~Carr and M.~Sasaki for discussions. V.DL., G.F. and 
A.R. are supported by the Swiss National Science Foundation 
(SNSF), project {\sl The Non-Gaussian Universe and Cosmological Symmetries}, project number: 200020-178787.
G.F. would like to thank the Instituto de Fisica Teorica (IFT UAM-CSIC) in Madrid for its support via the Centro de Excelencia Severo Ochoa Program under Grant SEV-2012-0249.
P.P. acknowledges financial support provided under the European Union's H2020 ERC, Starting 
Grant agreement no.~DarkGRA--757480, under the MIUR PRIN and FARE programmes (GW-NEXT, CUP:~B84I20000100001), and 
support from the Amaldi Research Center funded by the MIUR program `Dipartimento di 
Eccellenza" (CUP:~B81I18001170001).

\bigskip


\end{document}